\documentclass[11pt]{article}
\usepackage{epsfig,amsfonts}
\usepackage[fleqn]{amsmath}
\usepackage{amsthm,amssymb}
\usepackage{graphicx}
\usepackage{hhline}
\usepackage{cite}

\def\be{\begin{equation}}
\def\ee{\end{equation}}
\def\bdm{\begin{displaymath}}
\def\edm{\end{displaymath}}
\def\bea{\begin{eqnarray}}
\def\eea{\end{eqnarray}}

\begin{document}

\begin{titlepage}
\begin{flushright}
  BONN\,-TH\,-2004-02\\
\end{flushright}

\vspace{1.5cm}

\begin{center}
\begin{LARGE}
  {\bf Expectation value of composite field $T{\bar T}$ in 
two-dimensional quantum field theory}

\end{LARGE}

\vspace{1.3cm}

\begin{large}

  {\bf Alexander B. Zamolodchikov}$^{*}$
\end{large}

\vspace{.5cm}

{Physikalisches Institut der Universit${\rm \ddot a}$t  Bonn\\
  Nu\ss allee 12, D-53115 Bonn, Germany}

\vspace{.2cm}

\vspace{.2cm}

{$^{*}$ On leave of absence from: NHETC,\\ 
Department of Physics and Astronomy\\
  Rutgers University\\
  Piscataway, NJ 08855-0849, USA}

\vspace{1.5cm}

\centerline{\bf Abstract} 

\vspace{.8cm} 

\parbox{11cm}

{ I show that the expectation value of the composite field $T{\bar T}$, 
  built from the components of the energy-momentum tensor, is expressed 
  exactly through the expectation value of the energy-momentum tensor 
  itself. The
  relation is derived in two-dimensional quantum field theory under broad 
  assumptions, and does not require integrability. }
\end{center}

\begin{flushleft}
  \rule{5.1 in}{.007 in}\\
  {January 2004}
\end{flushleft}
\vfil

\end{titlepage}
\newpage

\section{Introduction}

Determination of one-point expectation values of local fields is an important
problem of quantum field theory (see e.g. \cite{PP}\cite{3A}). 
The expectation values $\langle {\cal O}_i
\rangle$ control linear reaction of the system to external forces which
couple to the fields ${\cal O}_i (z)$\,. Also, in view of the operator-product
expansions (OPE)
\bea\label{ope} {\cal O}_i (z)\,{\cal O}_j (z') =
\sum_{k}\,C_{i\,j}^{k} (z-z')\,{\cal O}_k (z')
\eea
the two-point correlation
functions $\langle {\cal O}_i (z) \,{\cal O}_j (z')\rangle$ (and, by repeated
application of \eqref{ope}, the multipoint correlation functions) are
expressed through the OPE structure functions $C_{ij}^{k}(z-z')$ and the
one-point expectation values $\langle {\cal O}_k \rangle$\,. But while the
structure functions (which describe local dynamics of the field theory)
usually admit perturbative expansions \cite{AlZ1}, the one-point expectation
values (incorporating information about the vacuum state of the theory) are
typically nonperturbative quantities \cite{PP}\cite{3A}\cite{AlZ1}\,, and no
general approach to their systematic evaluation is known \footnote{In two
  dimensions, rather accurate numerical estimates can be obtained in many cases
  through some version of the Truncated Conformal Space Approach, 
  see Ref. \cite{guida}}\,.

In recent years, some progress was made in determination of the
one-point expectation values in 2D integrable quantum field
theories \cite{LZ1}\cite{FLZZ1}\cite{FLZZ2}\cite{X}\cite{XX}\cite{pascal}\,. 
In particular,
in Ref.\cite{FLZZ2} exact expectation values of the lowest
nontrivial descendant fields were obtained in the cases of the
sine-Gordon model and the $\Phi_{(1,3)}$-perturbed minimal CFT.
Simplest of these descendants is the composite field $T{\bar T}$
built from the chiral components $T$, ${\bar T}$ of the
energy-momentum tensor $T_{\mu\nu}$\,. Remarkably, the result of
\cite{FLZZ2} shows that in these cases the expectation value of
$T{\bar T}$ relates to the expectation value of the trace
component $\Theta = \frac{\pi}{2}\,T_{\mu}^{\mu}$ as follows
\bea\label{basic1} \langle\,T{\bar T}\,\rangle = -\,
\langle\,\Theta\,\rangle^2\,. \eea 
Subsequently, expectation
values of the lowest descendant fields were obtained in few other
integrable models, including the Bullough-Dodd model and the
$\Phi_{(1,2)}$-perturbations of the minimal CFT \cite{pascal}, and
again in all these cases the expectation values of $T{\bar T}$ and
$\Theta$ turned out to fulfill the relation \eqref{basic1}\,.

In this note I will show that the relation \eqref{basic1} (and indeed somewhat
more general relation \eqref{basic2} below) holds in 2D quantum field theory
under rather broad assumptions; in particular, the theory {\it is not}
required to be integrable.

In the Refs\cite{FLZZ2} and \cite{pascal} the field theories were assumed to
live on an infinite Euclidean plane. One can consider instead a field 
theory on an infinite cylinder, with one of the Euclidean axis compactified 
on a circle (this of course is the Matsubara representation of the 
field theory at finite temperature). I will show that in this case the 
relation \eqref{basic1} generalizes as follows, 
\bea\label{basic2} \langle\,T{\bar
  T}\,\rangle = \langle\, T \,\rangle \langle\,{\bar T}\,\rangle -
\langle\,\Theta\,\rangle \langle \,\Theta\,\rangle\,.  
\eea 
When
the circumference of the cylinder goes to infinity (equivalently,
the temperature goes to zero) the global rotational symmetry is
restored, making the expectation values of the chiral components
$T$ and ${\bar T}$ vanish - in this limit \eqref{basic2} reduces
to \eqref{basic1}\,.

I will also argue that the relation \eqref{basic2} remains valid if the
vacuum expectation
values $\langle\, ...\, \rangle = \langle\, 0 \mid ... \mid 0\, \rangle$ there
are replaced by more general diagonal matrix elements 
$\langle\, n \mid ... \mid n\, \rangle$, where $\mid n \,\rangle$ is any 
non-degenerate eigenstate of the energy and momentum operators
(in the case of cylinder, to make this statement precise one has to
take the Hamiltonian picture in which the coordinate along the cylinder is
taken as the Euclidean time).

I present the arguments in sections 2 through 4 below. In section 5 I give 
another form of the relation \eqref{basic2}, which can be useful in analysis
of critical singularities in 2D statistical systems, and in other
applications. 

Throughout this paper I consider quantum field theory in flat 2D
space, and my discussion is in terms of the Euclidean version of the theory.
The points $z$ of the 2D space can be labeled by the Cartesian coordinates
$({\rm x}, {\rm y})$, but I will usually use complex coordinates
$z = ({\rm z},{\bar{\rm z}}) = ({\rm x} + i{\rm y}, {\rm x} - i{\rm y})$.
I assume usual normalization of the energy-momentum tensor $T_{\mu\nu}$: 
for instance, in
the picture where ${\rm y}$ is taken as the Euclidean time, $- T_{\rm yy}$
coincides with the energy density. The chiral components $T$, ${\bar T}$,
$\Theta$ are normalized
according to the CFT convention \cite{CFT}, namely $T = -(2\pi)\,T_{\rm z z}$, 
${\bar T} = -(2\pi)\,T_{{\bar{\rm z}}{\bar{\rm z}}}$, $\Theta =
(2\pi)\,T_{{\rm z}{\bar{\rm z}}}$. 

\section{Assumptions and sketch of the argument}

In this section I list basic assumptions about the field theory
and display main idea of my arguments. More subtle points,
including precise definition of the field $T{\bar T}$, are
discussed in the next two sections. Some of the assumptions
concern with the local dynamics of the field theory, others are
about the global settings (the geometry of the space). I will
stress the distinction by giving them additional labels (L) or
(G).

My basic assumptions are as follows:

\vspace{.2cm}

1 (L). {\it Local translational and rotational symmetry}. This
implies existence of local field $T_{\mu\nu}$ (the energy-momentum
tensor) which is symmetric, $T^{\mu\nu}(z) = T^{\nu\mu}(z)$, and
satisfies the continuity equation $\partial_{\mu} T^{\mu\nu}(z)
=0$. In terms of the conventional chiral components $T = - 2\pi\,
T_{\rm
zz}$, ${\bar T} = - 2\pi\,T_{{\bar{\rm z}}{\bar{\rm z}}}$ and $\Theta =
2\pi\,T_{{\rm z}{\bar{\rm z}}} = 2\pi\,T_{{\bar{\rm z}}{\rm z}}$ the
continuity equation is written as 
\bea\label{continuity}
&&\partial_{{\bar{\rm z}}} T({z}) =
\partial_{{\rm z}}\Theta({z})\,,\\
&&\partial_{{\rm z}} {\bar T}({z}) =
\partial_{{\bar{\rm z}}} \Theta({z})\,.
\eea 
This assumption is already taken into account in writing the
OPE \eqref{ope}\,, where the structure functions $C_{ij}^{k}$ are
assumed to depend on the separations $z-z'$ only.

\vspace{.2cm}

2 (G). {\it Global translational symmetry}. I assume that
for any local field ${\cal O}_i (z)$ the expectation value
$\langle\,{\cal O}_i (z) \,\rangle$ is a constant independent of $z$.
It follows from \eqref{ope} that the two-point correlation functions
depend only on the separations,
\bea\label{twopoint}
\langle\,{\cal O}_i (z) {\cal O}_j (z')\,\rangle = G_{ij}(z-z')\,.
\eea

\vspace{.2cm}

3 (G). {\it Infinite separations}. I assume that at least one direction
(i.e. Euclidean vector
$e = ({\rm e}, {\bar{\rm e}})$) exists, such that for any ${\cal O}_i$
and ${\cal O}_j$
\bea\label{ass}
\lim_{t\to\infty}\,\langle\,{\cal O}_i (z+e\,t) {\cal O}_j (z')\,\rangle
= \langle\,{\cal O}_i\,\rangle \langle \,{\cal O}_j \,\rangle\,.
\eea

The ``global'' assumptions 2 and 3 imply that the underlying geometry of
2D space is either an infinite plane, or an infinitely long cylinder.

\vspace{.2cm}

4. (L) {\it CFT limit at short distances}. I will assume that the
short-distance behavior of the field theory is governed by a
conformal field theory, and that certain no-resonance condition is
satisfied. I will detail the content of this assumption in section
4 below. Here I just mention that this assumption is needed in order 
to make definition of the composite field $T{\bar T}$ essentially 
unambiguous \footnote{There is intrinsic ambiguity 
  in adding certain total derivatives, which does not
  affect the expectation value $\langle T{\bar T} \rangle$. I discuss 
  it in section 4.}.

\vspace{.2in}

The main idea of my arguments stems from simple identity
involving two-point correlation functions of the energy-momentum
tensor, consequence of the assumptions 1-3 alone. Consider the
following combination of two-point correlation functions
\bea\label{cdef} {\cal C} = \langle\,T(z){\bar T}(z')\,\rangle -
\langle\,\Theta(z) \Theta(z')\,\rangle\,. 
\eea 
Take
$\partial_{\bar{\rm z}}$ derivative of \eqref{cdef} and transform
it as follows. In the first term in \eqref{cdef}, use the equation
(4) to replace the derivative $\partial_{\bar{\rm z}}T(z)$ by
$\partial_{\rm z}\Theta(z)$, and then apply the
Eq.\eqref{twopoint} to move the derivative to the second entry
${\bar T}(z')$. When the derivative $\partial_{\bar{\rm z}}
\Theta(z)$ in the second term is also moved to $\Theta(z')$, one
finds 
\bea\label{transform} \langle\,\partial_{\bar{\rm
z}}T(z)\,{\bar T}(z') -
\partial_{\bar {\rm z}}\Theta(z)\,\Theta(z')\,\rangle = \langle
\,-\Theta(z)\,\partial_{{\rm z}'}{\bar T}(z') +
\Theta(z)\,\partial_{\bar{\rm z}'}\Theta(z')\,\rangle = 0\,, 
\eea
where the equation (5) was used in the last step. By similar
transformations one shows that the $\partial_{\rm z}$ derivative
of \eqref{cdef} also vanishes, and hence the combination
\eqref{cdef} is a constant, independent of the coordinates.

Note that in this derivation only the first two assumptions 1 and
2 are used. Adding the assumption 3 allows one to relate this
constant to the one-point expectation values of the fields
involved. Taking the limit \eqref{ass} of the right-hand side of
Eq.\eqref{cdef} one finds 
\bea\label{basic3} {\cal C} =
\langle\,T\,\rangle\langle\,{\bar T}\,\rangle -
\langle\,\Theta\,\rangle\langle\,\Theta\,\rangle\,. 
\eea

On the other hand, some meditation about the equation \eqref{cdef}
makes it plausible that the constant ${\cal C}$ also coincides
with the expectation value of appropriately defined composite
operator $T{\bar T}$. Indeed, one expects that the composite field
$T{\bar T}$ can be obtained in some way from the product $T(z){\bar
T}(z')$ by bringing the points $z$ and $z'$ together. The main
obstacle is in the presence of singular terms in the
operator product expansion of $T(z){\bar T}(z')$, which make
straightforward limit impossible. As we will see in the next
section, the second term in the combination $T(z){\bar T}(z') -
\Theta(z)\Theta(z')$ exactly subtracts these singular terms, so
that the limit $z\to z'$ in \eqref{cdef} can be taken, leading to
\eqref{basic2}\,.

\section{Operator product expansions}

It is not difficult to repeat manipulations of the previous
section, this time working not with the two-point functions
\eqref{cdef} but with the combination of the operator products
$T(z){\bar T}(z') - \Theta(z)\Theta(z')$ itself. Using only (4)
and (5), one finds 
\bea\label{opeid1}
\partial_{{\bar z}}\big(T(z){\bar T}(z') -
\Theta(z)\Theta(z')\big) = &\ \ \ \ \ \ \ \ \ \ \ \ \ \ \nonumber
\\ \big(\partial_{\rm z}+
\partial_{{\rm z}'}\big)\Theta(z){\bar T}(z')& - \ \
\big(\partial_{\bar{\rm z}} +
\partial_{{\bar{\rm z}}'}\big)\Theta(z)\Theta(z')\,,
\eea and \bea\label{opeid2}
\partial_{{z}}\big(T(z){\bar T}(z') - \Theta(z)\Theta(z')\big)
=&\ \ \ \ \ \ \ \ \ \ \ \ \ \nonumber\\ \big(\partial_{\rm z} +
\partial_{{\rm z}'}\big)T(z){\bar T}(z')& - \ \
\big(\partial_{\bar{\rm z}} +
\partial_{{\bar{\rm z}}'}\big)T(z)\Theta(z')\,.
\eea 
The meaning of these equations becomes clearer after
inserting the operator product expansions 
\bea\label{tope1}
\Theta(z){\bar
T}(z') = &\sum_{i}\,B_i (z-z')\,{\cal O}_i (z')\,,\\
T(z)\Theta(z') = &\sum_{i}\,A_i (z-z')\,{\cal O}_i (z')\,, 
\eea
and 
\bea\label{tope}
T(z){\bar T}(z') = &\sum_{i} \,D_i (z-z')\,{\cal O}_i (z')\,,
\\
\Theta(z)\Theta(z') = &\sum_{i} \,C_i (z-z')\,{\cal O}_i (z')\,,
\eea 
where the sums involve complete set of local fields $\{{\cal
O}_i\}$\,. The equations \eqref{opeid1}, \eqref{opeid2} then read
\bea\label{opeidd1} \sum_{i}\,\partial_{\bar{\rm
z}}F_{i}(z-z')\,{\cal O}_i (z')= &\nonumber
\\
\sum_{i}\bigg(B_i (z-z')\,\partial_{{\rm z}'}{\cal O}_i (z')& - \
\ C_i (z-z')\,\partial_{{\bar{\rm z}}'}{\cal O}_i (z')\bigg)\,,
\eea \bea\label{opeidd2} \sum_{i}\,\partial_{{\rm
z}}F_{i}(z-z')\,{\cal O}_i (z')= &\nonumber
\\
\sum_{i}\bigg(D_i (z-z')\,\partial_{{\rm z}'}{\cal O}_i (z')& - \
\ A_i (z-z')\,\partial_{{\bar{\rm z}}'}{\cal O}_i (z')\bigg)\,,
\eea where \bea\label{deltadef} F_i (z-z') = D_i (z-z') - C_i
(z-z')\,.\eea
Note that the right-hand sides of the Eq's \eqref{opeidd1},
\eqref{opeidd2} involve only coordinate derivatives of local
fields. It follows that any operator ${\cal O}_i$ appearing in the
expansion 
\bea\label{diffope1} T(z){\bar T}(z') -
\Theta(z)\Theta(z') = \sum_i \,F_i (z-z')\,{\cal O}_i (z')\,, 
\eea
unless itself is a coordinate derivative of another local
operator, comes with a constant (i.e. coordinate-independent)
coefficient $F_i$. In other words, the operator product expansion
\eqref{diffope1} can be written as 
\bea\label{diffope2} T(z){\bar
T}(z') - \Theta(z)\Theta(z') = {\cal O}_{T{\bar T}} (z') + {\rm
derivative\ terms}\,, 
\eea 
where ${\cal O}_{T{\bar T}}(z)$ is some
local operator. At this point it is possible to {\it define} the
composite field $T{\bar T}$ through the Eq.\eqref{diffope2}:
\bea\label{ttdef}
T{\bar T}(z): = {\cal O}_{T{\bar T}}(z)\,;
\eea 
then the desired
relation \eqref{basic2} follows immediately. Note that although 
in this way 
one defines $T{\bar T}$ only modulo derivative terms, in view of the
assumption 2 those terms 
bring no contribution to the left-hand side of
\eqref{basic2}\,. However, this definition may look a bit too formal
to bring much insight into the meaning of \eqref{basic2}\,. 
To understand the nature of the 
limit $z \to z'$ in Eq.\eqref{diffope1}\,, and thus to make 
contact with more constructive definition of the composite field $T{\bar T}$, 
one needs to know more about short-distance behavior of the 
field theory. In the the present discussion, this 
information is furnished through the assumption 4 
(see section 2). Let me now describe its content and 
implications.  

\section{Dimensional analysis}

As was mentioned in section 2, I assume that the 
short-distance limit of the field theory is controlled by certain 
conformal field theory, which I will refer to simply as the CFT. 
More precisely, I
will assume that the field theory at hand is the CFT perturbed by its
relevant operators. To avoid unnecessarily complex expressions, let me first
assume that the perturbation is by a single operator $\Phi_{\Delta}$ of the
dimensions $(\Delta, \Delta)$ with $\Delta < 1$; then the theory is 
described by the action
\bea\label{pcft1}
{\cal A} = {\cal A}_{CFT} + \mu\,\int\,\Phi_{\Delta}(z) \,d^2 z\,,
\eea
where $\mu$ is a coupling constant which has the dimension $[{\rm
  length}]^{2\Delta - 2}$. This formulation of the theory makes it possible to
carry out dimensional analysis of the structure functions in
\eqref{tope}. 

Let $\{{\cal O}_i\}$ be complete set
of local fields of the CFT, including primary fields as well as 
their descendents, and let $(\Delta_i , {\bar\Delta}_i )$ be the left and
right scale dimensions of the fields ${\cal O}_i$. This set includes the
field $T{\bar T}$ (of the dimensions $(2,2)$), which in CFT is just 
the descendant $T{\bar T} = L_{-2}
{\bar L}_{-2} I$ of the identity operator. Equivalently, this field can be
defined as $T{\bar T}(z') = {\rm lim}_{z\to z'} T(z){\bar T}(z')$, where the
limit is straightforward since in CFT the above operator product has no
singularity at $z=z'$. 

As was explained in Ref.\cite{mee}, the fields ${\cal O}_i$ of the perturbed 
theory \eqref{pcft1} are in one-to-one correspondence with the fields of the
CFT (hence I use the same notations). The field ${\cal O}_i$ has the
spin $s_i = \Delta_i - {\bar\Delta}_i$ and the mass dimension $d_i =
\Delta_i + {\bar\Delta}_i$, and ${\cal O}_i$ coincides with the corresponding
CFT field in the limit $\mu \to 0$. Unless certain resonance conditions are 
met (see below), these properties characterize the field ${\cal O}_i$ 
uniquely \cite{mee} (see also \cite{FLZZ2}). One says that the field 
${\cal O}_i$ has $n$-th order resonance with the field ${\cal O}_j$ if these
fields have the same spins, $s_j = s_i$, and their dimensions satisfy the
equation $d_i = d_j + 2n\,(1-\Delta)$ (the resonance condition) with some 
positive integer $n$. When this resonance condition is fulfilled the above 
characterization of the field ${\cal O}_i$ allows for the ambiguity ${\cal
  O}_i \to {\cal O}_i + {Const}\,\,\mu^n\,{\cal O}_j$.

The field $T{\bar T}$ always has intrinsic ambiguity of
the form $T{\bar T} \to T{\bar T} + {Const}\,\,\partial_{\rm
  z}\partial_{\bar{\rm z}}\Theta$, where $\Theta$ is the trace component of
the energy-momentum tensor of the perturbed theory. Since in \eqref{pcft1}
$\Theta = (1-\Delta)\,\pi \mu\,\Phi_{\Delta}$, the ambiguity is due to the
first-order resonance of $T{\bar T}$ with the derivative $\partial_{\rm z}
\partial_{\bar{\rm z}}\Phi_{\Delta}$. However, this ambiguity has no 
effect on the expectation value of $T{\bar T}$. For the present analysis 
the danger is 
in possible resonances with non-derivative fields. Since at this time I do not
know how to handle the resonance cases, I accept the
following no-resonance assumption:

\vspace{.2cm}

4'. \  Dimensions $\Delta_i$ of the fields ${\cal O}_i$ of the CFT
satisfy the condition
\bea\label{nores1}
\Delta_i - 2 + n\,(1 - \Delta) \neq 0 \qquad {\rm for} \quad n=1,2,3, ...\,,
\eea
with the only exception of 
$\Delta_i = \Delta + 1$ (which is the dimension of $\partial_{\rm z}
\partial_{\bar{\rm z}}\Phi_{\Delta}$).

\vspace{.2cm}

According to \cite{AlZ1}, the OPE structure functions in \eqref{ope} 
admit power-series expansions in $\mu$, with the coefficients computable 
(in principle) through the conformal perturbation theory. Thus, the structure
functions $D_i (z-z')$ in \eqref{tope} can be written as
\bea\label{dexpan}
D_i (z-z') = \sum_{n=0}^{\infty}\,({\rm z} - {\rm z}')^{\Delta_i - 2 +
  n\,(1-\Delta)}\,({\bar{\rm z}}-{\bar{\rm z}}')^{{\bar\Delta}_i - 2 +
  n\,(1-\Delta)} \, D_{i}^{(n)}\,\mu^{n}\,.
\eea
The zero-order coefficients $D_{i}^{(0)}$ are
taken from the unperturbed CFT, hence $D_{i}^{(0)} = 0$ unless ${\cal O}_i$
is the field $T{\bar T}$ or one of its derivatives, and $D_{T{\bar T}}^{(0)}
= 1$. Then it follows from \eqref{nores1} that the only terms in the 
expansions \eqref{dexpan} which carry vanishing powers of both
${\rm z} - {\rm z}'$ and ${\bar {\rm z}}-{\bar{\rm z}}'$ are the 
zero-order term of $D_{T{\bar T}}$, and the first-order term associated 
with 
${\cal O}_i = \partial_{{\rm z}'}\partial_{{\bar{\rm z}}'}\Phi_{\Delta}$.  

Similar expansion can be written down for the structure functions $C_i (z-z')$
in the OPE (16),
\bea\label{cexpan}
C_i (z-z') = \sum_{n=2}^{\infty}\,({\rm z} - {\rm z}')^{\Delta_i - 2 +
  n\,(1-\Delta)}\,({\bar{\rm z}}-{\bar{\rm z}}')^{{\bar\Delta}_i - 2 +
  n\,(1-\Delta)} \, C_{i}^{(n)}\,\mu^{n}\,.
\eea
Note that the sum here starts from
$n=2$, consequence of the fact that $\Theta \sim \mu\,\Phi_{\Delta}$. 
In this case the no-resonance condition \eqref{nores1} implies that 
there are no terms
with vanishing powers of both ${\rm z} - {\rm z}'$ and 
${\bar {\rm z}}-{\bar{\rm z}}'$ at all.

Consider now the differences $F_i (z-z') = D_i (z-z') - C_i (z-z')$. It
 follows from \eqref{opeidd1}, \eqref{opeidd2} that, unless ${\cal O}_i$ is a
 derivative of another local field, all terms with nonzero powers of 
${\rm z} - {\rm z}'$ or ${\bar {\rm z}}-{\bar{\rm z}}'$ must cancel out 
in this difference \footnote{This implies for instance
$D_{T{\bar T}}^{(1)} =0$, the statement which is easily verified in conformal
 perturbation theory.}. Therefore 
\bea\label{fjs}
F_i (z-z') = 0 \qquad {\rm unless} \quad {\cal O}_i = T{\bar T} \quad {\rm or} 
\quad {\cal O}_i = {\rm derivative}\,,
\eea
and
\bea\label{fjtt}
F_{T{\bar T}} (z-z') = 1\,.
\eea
One concludes that the definition of $T{\bar T}$ through the conformal
 perturbation theory agrees with the formal definition \eqref{ttdef}\,.

It is not difficult to generalize this analysis to the case when the
CFT is perturbed by a mixture
$\sum_{a}\,\mu_a\,\int\,\Phi_{\Delta_a}(z)\,d^2 z$ of relevant operators
$\Phi_{\Delta_a}$. The dimensional analysis can be carried out in a similar 
straightforward way provided the no-resonance condition is modified as follows:

\vspace{.2cm}

4''.\ Dimensions $\Delta_i$ of the fields ${\cal O}_i$ of the CFT
satisfy the conditions
\bea\label{nores2}
\Delta_i - 2 + \sum_a \,n_a\,(1 - \Delta_a) \neq 0 
\eea
for any non-negative integers $n_a$ such that $\sum_a\,n_a > 0$, with the 
only exceptions of $\Delta_i = \Delta_a + 1$.

\vspace{.2cm}

\section{Further remarks}

Let the 2D space be a cylinder, with one of the Cartesian coordinates
compactified on a circle of circumference $R$, $({\rm x}, {\rm y}) \sim 
({\rm x}+R, {\rm y})$, and let ${\mathbb H}$ and ${\mathbb P}$ be the 
Hamiltonian and the momentum 
operators in the picture where the coordinate ${\rm y}$ along the cylinder is
taken as the Euclidean time. The arguments of the previous sections validate
the relation \eqref{basic2} with $\langle\, ... \,\rangle$ standing for 
the matrix element $\langle \,0 \mid \, ... \,\mid 0\,\rangle$, where
$\mid 0\,\rangle$ is the ground state of the Hamiltonian ${\mathbb H}$ 
(and I assume
that $\langle\,0\mid 0\,\rangle =1$). But it is not
difficult to show that the same relation \eqref{basic2} remains valid if the
vacuum expectation values there are replaced by generic diagonal matrix 
elements $\langle\,n\mid \, ... \,\mid n\,\rangle$, where $\mid n\,\rangle$ 
is an arbitrary non-degenerate eigenstate of the energy and momentum operators,
\bea\label{eigenstate}
{\mathbb H} \mid\,n\,\rangle = E_n \mid\,n\,\rangle\,, \qquad 
{\mathbb P} \mid\,n\,\rangle = P_n \mid\,n\,\rangle\,,
\eea
and again the normalization $\langle\,n\mid n\,\rangle =1$ is assumed.
Indeed, of the assumptions listed in section 2, the local ones (1 and 4) are
independent on the choice of matrix element, while the assumption 2 (global
translational invariance) certainly remains valid when any diagonal matrix
element between energy-momentum eigenstates is taken. Hence one can repeat
the calculation at the end of section 2 (which uses only the assumptions 1 and
2) and show that again the combination
\bea\label{cndef}
{\cal C}(n) = \langle\,n \mid \,T(z){\bar T}(z')\,\mid n\,\rangle - 
\langle\,n\mid\,\Theta(z)\Theta(z')\,\mid n
\,\rangle 
\eea
is a constant, independent of the points $z$ and $z'$. In general, the 
asymptotic factorization \eqref{ass} no longer holds, since the two-point 
function \break
\hbox{$\langle \,n\mid\,{\cal O}_i ({\rm x},{\rm y}) {\cal O}_j ({\rm x}', {\rm
  y}') \,\mid n \,\rangle$} can pick up contributions from the intermediate
states $\mid n'\,\rangle$ with $E_{n'} < E_n$ which give rise to terms growing
exponentially with $|{\rm y-y'}|$. However, one can write down the spectral
decompositions of the two-point functions in the right-hand side of
\eqref{cndef}, i.e.
\bea\label{spectral}
\langle \,n\mid T(z){\bar T}(z')\mid n \,\rangle =
\sum_{n'}&\langle\,n\mid T(z)\mid n' \,\rangle \langle\,n'\mid {\bar T}(z)
\mid n\,\rangle \times\nonumber\\
&e^{(E_n - E_{n'})|{\rm y-y'}| + i(P_n - P_{n'})({\rm x-x'})}
\eea
and similar decomposition of $\langle\,n\mid\Theta(z)\Theta(z')\mid
n\,\rangle$, where $({\rm x, y})$ and $({\rm x', y'})$ are Cartesian
coordinates of the points $z$ and $z'$, respectively. Clearly, for the
combination \eqref{cndef} to be independent of the coordinates, all 
terms in these decompositions with $n' \neq n$ must cancel out 
between the two correlators in
the right-hand side of \eqref{cndef}\,. If $\mid n \,\rangle$ is
non-degenerate, it follows that
\bea\label{cnfactor}
{\cal C}(n) = \langle\,n\mid T(z)\mid n \,\rangle\langle\,n \mid {\bar
  T}(z')\mid n\,\rangle - 
\langle\,n\mid \Theta(z)\mid n \,\rangle\langle\,n \mid \Theta(z')
\mid n\,\rangle\,,
\eea
and by taking the limit $z \to z'$ one arrives at the desired relation
\bea\label{basic4}
\langle\, n \mid T{\bar T} \mid n \,\rangle = 
\langle\,n\mid T \mid n \,\rangle\langle\,n \mid {\bar T}\mid n\,\rangle 
- \langle\,n\mid \Theta \mid n \,\rangle\langle\,n \mid \Theta
\mid n\,\rangle\,.
\eea

It is useful to rewrite the relation \eqref{basic4} in
somewhat different form. In terms of Cartesian components of the
energy-momentum tensor $T_{\mu\nu}$ the Eq.\eqref{basic4} reads \footnote{The
  factor $\pi^2$ here is due to the factor $2\pi$ in the 
  definition of the
  chiral components $T$ and ${\bar T}$, see sect.1}
\bea\label{cncartes}
(34) = -\pi^2\,\big(\langle\,n\mid T_{\rm yy}\mid n\,\rangle\langle 
\,n \mid T_{\rm xx}\mid n \,\rangle - 
\langle\,n\mid T_{\rm xy}\mid n\,\rangle\langle 
\,n \mid T_{\rm xy}\mid n \,\rangle\big),
\eea
On the other hand, by the very meaning of the energy-momentum tensor we have 
\bea\label{texp1}
\langle\,n\mid T_{\rm yy}\mid n \,\rangle = - \frac{1}{R}\,E_{n}(R)\,,\qquad 
\langle\,n\mid T_{\rm xx}\mid n \,\rangle = - \frac{d}{dR}\,E_{n}(R)\,,
\eea
and
\bea\label{texp2}
\langle\,n\mid T_{\rm xy}\mid n \,\rangle = \frac{i}{R}\,P_{n}(R)\,.
\eea
where I have indicated explicitly the $R$-dependence of the energy-momentum
eigenvalues (of course, the $R$-dependence of $P_n$ is fixed by the momentum
quantization condition: $P_n (R) = 2\pi\,p_n /R$, where $p_n$ are
$R$-independent integers). Thus the expectation value \eqref{basic4} can be
expressed in terms of the eigenvalues $E_n (R), P_n (R)$ as follows
\bea\label{basic5}
\langle \,n\mid T{\bar T}\mid n\,\rangle = - \frac{\pi^2}{R}\,
\bigg(E_{n}(R)\frac{d}{dR} E_{n}(R) + \frac{1}{R}\,P_{n}^2 (R)\bigg)\,.
\eea

Suppose the field theory \eqref{pcft1} is massive, with $M_0$ being 
the mass of its lightest particle. Then for $R>>M_{0}^{-1}$ the 
ground-state energy $E_0(R)$ approaches its asymptotic linear form with 
exponential accuracy, i.e.
\bea\label{venergy}
E_0 (R) = F_0\,R + O\big( e^{-M_0 R}\big)\,,
\eea
where $F_0$ is the vacuum energy density in infinite space. In the same limit, 
the first excited state $\mid 1 \,\rangle$ corresponds to the one-particle
state with zero momentum, hence
\bea\label{menergy}
E_1 (R) = F_0\,R + M_0 +  O\big( e^{-M_0 R}\big)\,.
\eea
Then it follows from \eqref{basic4} that (up to terms $\sim e^{-M_0 R}$)
\bea\label{lowexp}
\frac{1}{\pi^2}\,\langle \,0 \mid T{\bar T}\mid 0\,\rangle = - F_{0}^2 \,, 
\qquad 
\frac{1}{\pi^2}\,\langle \,1 \mid T{\bar T}\mid 1\,\rangle = -  F_{0}^2 -
\frac{1}{R}\,F_{0}\,M_{0}\,.
\eea
These expressions can be useful in analysis of subleading singularities 
in statistical systems near criticality, in the situations 
where the irrelevant operator $T{\bar T}$ plays significant role. 
This is the case, for instance, for the Ising phase transition near 
the Ising tri-critical point, because the RG flow from the tricritical fixed
point (the $c=7/10$ minimal CFT) down to the Ising fixed point 
(the $c=1/2$ minimal CFT) arrives at the latter along direction which
contains the field $T{\bar T}$ as its most significant (i.e. least irrelevant)
component \cite{AlZ2}\,. Another example is the Ising field theory 
with pure imaginary
magnetic field, taken near the Yang-Lee singularity. In such cases the 
relations \eqref{lowexp} lead to predictions about amplitudes of subleading
singular terms in the expansions of the free energy and correlation length
near the critical point. I intend to discuss these applications elsewhere.

\bigskip

\section*{Acknowledgments}

\noindent
I am grateful to S.Lukyanov, Al.Zamolodchikov, A.Polyakov, D.Kutasov, and 
G. von Gehlen for discussions and interest to this work.
This research is supported in part by DOE grant $\#$DE-FG02-96 ER 40949, and
by Alexander von Humboldt Foundation.

\end{document}